# Attenuation Correction of L-shell X-ray Fluorescence Computed Tomography Imaging[*]

Liu Long(刘珑)[1,2], Huang Yang(黄旸)[1,2], Ma Bo(马波)[1,2], Xu Qing(徐清)[1], Yan Lingtong(闫灵通)[1], Li Li(李丽)[1], Feng Songlin(冯松林)[1], Feng Xiangqian(冯向前)[1;1)]

[1] Key Laboratory of Nuclear Radiation and Nuclear Energy Technology, Institute of High Energy Physics, Chinese Academy of Sciences, Beijing 100049, China

[2] University of Chinese Academy of Sciences, Beijing 100049, China

**Abstract**: X-ray Fluorescence Computed Tomography(XFCT) is a prevalent experimental technique which is utilized to investigate the spatial distribution of elements in sample. The sensitivity of L-shell XFCT of some elements is lower than that of K-shell XFCT. However, the image reconstruction for this technique has much more problems than that of transmission tomography, one of which is self-absorption. In the present work, a novel strategy was developed to deal with such problems. But few researches are concerned on attenuation correction of L-shell XFCT that are essential to get accurate reconstructed image. We make use of the known quantities and the unknown elemental concentration of interest to express the unknown attenuation maps. And then the attenuation maps are added in the contribution value of the pixel in MLEM reconstruction method. Results indicate that the relative error is less than 14.1%, which is proved this method can correct L-shell XFCT very well.

## 1 Introduction

X-ray fluorescence computed tomography (XFCT) is an experimental technique that can reconstruct the distribution of elements within the sample from the measurement of fluorescence stimulated from the sample[1]. The sample is irradiated with a x-ray beam, of which the energy is greater than the K-shell energy or L-shell energy of the interest elements. These x rays undergo photoelectric interaction and stimulate fluorescence of atoms. An energy-discriminating detector is placed at 90-deg. to detect the fluorescence undergoing attenuation for minimizing the effect by Compton Scattering[2]. Each element has its own fluorescence, and intensity of fluorescence can reflect content of element. The distribution of interest elements in sample can be reconstructed with measurement of fluorescence when the sample is scanned and rotated if attenuation can be neglected[3]. To reconstruct more accurate element distribution image, attenuation correction is necessary[4, 5]. Hogan presents a method about FBP with attenuation correction in the XFCT reconstruction, in which attenuation coefficient distribution at incident energy and fluorescence energy must be known[6]. Then Golosio comes up with a method that solves the attenuation problem combining x-ray fluorescence, Compton and transmission tomography[7].

La Riviere develops an alternating-update iterative reconstruction algorithm based on maximizing a penalized Poisson likelihood objective function[8]. In this work, the unknown linear attenuation coefficients at fluorescence energy are expressed as a linear combination of known attenuation coefficient at the incident energy and the unknown concentration of the element. However, this absorption method is only concerned to two energy regimes: that above the K-edge energy and that between K-edge energy and $L_1$-edge energy. Magdalena Bazalova researched the method of getting Pt distribution of Cisplatin with K-shell and L-shell XFCT[9]. Through the research, they found that the sensitivity of the K-shell XFCT with 80 keV was 4.4 and 3.0 times lowers than that of L-shell XFCT with 15 keV excitation beam for 2 cm and 4 cm diameter phantom. Cisplatin concentration error decreased form 63% to 12% when attenuation correction was incorporated in L-shell XFCT iterative reconstruction algorithm. However, few studies have been focused on attenuation correction of L-shell XFCT. In our work, the energy above M1-edge energy are interested and the attenuation coefficient at the incident energy are expressed as a linear combination of $E^{-2.83}$ and $E^{-2.665}$. The coefficient can be got by doul-energy method[10]. Then the unknown attenuation coefficient can be expressed as a function with unknown concentration ρ of the element. After that, the attenuation coefficients are added in the contribution value of the pixel in MLEM reconstruction method. We aim to test the feasibility of this method, and it is hoped that the attenuation correction of L-shell XFCT will be resolved with our proposed method.

[*] Supported by National Natural Science Foundation of China (11205167, 11305183, 10975147 and 11175190).
[1)] E-mail: fengxq@ihep.ac.cn

## 2 Method

### 2.1 Unknown attenuation maps

In order to obtain the unknown fluorescence maps at the energy $E_{K_\alpha}^{(n)}$ or $E_{L_1}^{(n)}$ of the element n of interest, the mass attenuation coefficient for a given element can be shown as (1)[11].

$$\left(\frac{\mu}{\rho}\right)^{(n)}(E) = C^{(n)}(E)E^{-\gamma^{(n)}(E)}. \tag{1}$$

Both $C^{(n)}(E)$ and $\gamma^{(n)}(E)$ are functions of energy but they change only when crossing absorption edges. In XFCT, two energy regimes are paid more attenuation for low-Z element: that above the K-edge energy $E_K^{(n)}$ and that between K-edge energy $E_K^{(n)}$ and $L_1$-edge energy $E_{L_1}^{(n)}$. And for high-Z element, four energy regimes are interest regions: that between K-edge energy $E_K^{(n)}$ and $L_1$-edge energy $E_{L_1}^{(n)}$, that between $L_1$-edge energy $E_{L_1}^{(n)}$ and $L_2$-edge energy $E_{L_2}^{(n)}$, that between $L_2$-edge energy $E_{L_2}^{(n)}$ and $L_3$-edge energy $E_{L_3}^{(n)}$ and that between $L_3$-edge energy $E_{L_3}^{(n)}$ and $M_1$-edge energy $E_{M_1}^{(n)}$.

$$C^{(n)}(E) = \begin{cases} C_1 & E > E_{K_\alpha} \\ C_2 & E_{K_\alpha} > E > E_{L_1} \\ C_3 & E_{L_1} > E > E_{L_2} \\ C_4 & E_{L_2} > E > E_{L_3} \\ C_5 & E_{L_3} > E > E_{M_1} \end{cases}. \qquad (2)$$

Where $C_1$, $C_2$, $C_3$, $C_4$ and $C_5$ are element-specific.

$$\gamma^{(n)}(E) = \begin{cases} 2.83 & E > E_{K_\alpha} \\ 2.6628 & E_{K_\alpha} > E > E_{L_1} \\ 2.6865 & E_{L_1} > E > E_{L_2} \\ 2.5825 & E_{L_2} > E > E_{L_3} \\ 2.5065 & E_{L_3} > E > E_{M_1} \end{cases}. \qquad (3)$$

Considering a mixture of N elements, which consist of $N_1$ elements those K-edge energy is lower than incident energy and the others $N_2$ elements those K-edge energy is higher than incident energy and $L_1$-edge energy is lower than incident energy. The liner attenuation coefficient at energy E can be written

$$\mu(E) = \sum_1^{N_1} C_1^{(n_1)}(E) E^{-2.83} \rho^{(n_1)} + \sum_1^{N_2} C_2^{(n_2)}(E) E^{-2.6628} \rho^{(n_2)}. \qquad (3)$$

which can be written as

$$\mu(E) = \left(\sum_1^{N_1} C_1^{(n)}(E) \rho^{(n_1)}\right) E^{-2.83} + \left(\sum_1^{N_2} C_2^{(n)}(E) \rho^{(n_2)}\right) E^{-2.6628}. \qquad (4)$$

If the K-edge energy of the highest-Z element N is lower than incident x-ray energy, The liner attenuation coefficient at energy $E_{K_\alpha}^{(N_1)}$ can be written

$$\mu\left(E_{K_\alpha}^{(N_1)}\right) =$$

$$\left(\sum_1^{N_1} C_1^{(n)}(E_I) \rho^{(n_1)}\right)\left(E_{K_\alpha}^{(N_1)}\right)^{-2.83} + \left(\sum_1^{N_2} C_2^{(n)}(E_I) \rho^{(n_2)}\right)\left(E_{K_\alpha}^{(N_1)}\right)^{-2.6628} +$$

$$\left(\left(E_{K_\alpha}^{(N_1)}\right)^{-2.6628} C_2^{(N_1)} - \left(E_{K_\alpha}^{(N_1)}\right)^{-2.83} C_1^{(N_1)}\right) \rho^{(N_1)}. \qquad (5)$$

If the $L_1$-edge energy of the highest-Z element N is lower than incident x-ray energy and the K-edge energy of that is higher than incident x-ray energy, The liner attenuation coefficient at energy $E_{L_1}^{(N_2)}$ can be written

$$\mu\left(E_{L_1}^{(N_2)}\right) =$$

$$\left(\sum_1^{N_1} C_1^{(n)}(E_I) \rho^{(n_1)}\right)\left(E_{L_1}^{(N_2)}\right)^{-2.83} + \left(\sum_1^{N_2} C_2^{(n)}(E_I) \rho^{(n_2)}\right)\left(E_{L_1}^{(N_2)}\right)^{-2.6628} +$$

$$\left(\left(E_{L_1}^{(N_2)}\right)^{-2.5065} C_3^{(N_2)} - \left(E_{L_1}^{(N_2)}\right)^{-2.6628} C_2^{(N_2)}\right) \rho^{(N_2)}. \qquad (6)$$

So the unknown attenuation coefficient $\mu\left(E_{K_\alpha}^{(N_1)}\right)$ at the energy $E_{K_\alpha}$ of the element $N_1$ and $\mu\left(E_{L_1}^{(N_2)}\right)$ at the energy $E_{L_1}$ of the element $N_2$ can be expressed as a function involving the known attenuation coefficient $\mu(E_I)$ and the unknown concentration ρ. Thus the question of image construction can be translated to the construction of the concentration.

After crossing the absorption edge of the highest-Z element, the reconstruction of the lower-Z element is possible with the concentration maps and attenuation maps of the higher-Z elements. The attenuation coefficient of the element $n_1$ of which the K-edge energy is lower than incident beam energy and the element $n_2$ of which the K-edge energy is higher than incident beam energy can be expressed by (7) and (8).

$$\mu(E_{K_\alpha}^{(n_1)}) = \left(\sum_1^{N_1} C_1^{(n)}(E_I)\rho^{(n_1)}\right)\left(E_{K_\alpha}^{(n_1)}\right)^{-2.83} + \left(\sum_1^{N_2} C_2^{(n)}(E_I)\rho^{(n_2)}\right)\left(E_{K_\alpha}^{(n_1)}\right)^{-2.6628} + \sum_{n_1+1}^{N_1}\left(\left(E_{K_\alpha}^{(n_1)}\right)^{-2.6628} C_2^{(i1)} - \left(E_{K_\alpha}^{(n_1)}\right)^{-2.83} C_1^{(i1)}\right)\rho^{(i1)} + \sum_{n_2+1}^{N_2}\left(\left(E_{K_\alpha}^{(n_1)}\right)^{-2.5065} C_3^{(i2)} - \left(E_{K_\alpha}^{(n_1)}\right)^{-2.6628} C_2^{(i2)}\right)\rho^{(i2)} + \left(\left(E_{K_\alpha}^{(n_1)}\right)^{-2.6628} C_2^{(n_1)} - \left(E_{K_\alpha}^{(n_1)}\right)^{-2.83} C_1^{(n_1)}\right)\rho^{(n_1)}. \quad (7)$$

$$\mu(E_{L_1}^{(n_2)}) = \left(\sum_1^{N_1} C_1^{(n)}(E_I)\rho^{(n_1)}\right)\left(E_{L_1}^{(n_2)}\right)^{-2.83} + \left(\sum_1^{N_2} C_2^{(n)}(E_I)\rho^{(n_2)}\right)\left(E_{L_1}^{(n_2)}\right)^{-2.6628} + \sum_{n_1+1}^{N_1}\left(\left(E_{L_1}^{(n_2)}\right)^{-2.6628} C_2^{(i1)} - \left(E_{L_1}^{(n_2)}\right)^{-2.83} C_1^{(i1)}\right)\rho^{(i1)} + \sum_{n_2+1}^{N_2}\left(\left(E_{L_1}^{(n_2)}\right)^{-2.5065} C_3^{(i2)} - \left(E_{L_1}^{(n_2)}\right)^{-2.6628} C_2^{(i2)}\right)\rho^{(i2)} + \left(\left(E_{L_1}^{(n_2)}\right)^{-2.5065} C_3^{(n_2)} - \left(E_{L_1}^{(n_2)}\right)^{-2.6628} C_2^{(n_2)}\right)\rho^{(n_2)}. \quad (8)$$

**2.2 Maximum Likelihood Expectation Maximization image reconstruction**

Image reconstruction methods include analytical method and iterative method. The typical method of analytical method is the Filtered Back Projection (FBP), and that of iterative method is the Maximum Likelihood Expectation Maximization (MLEM)[12]. The main advantage of FBP method is the fast reconstruction speed, and the main disadvantage is that the ability of anti-noise is poor. On the contrary, MLEM method has slower reconstruction speed and better anti-noise ability. Maximum Likelihood Expectation Maximization (MLEM) algorithm can be expressed as (9)[13].

$$f^{(p+1)}(i,j) = \frac{f^{(p)}(i,j)}{\sum_{m,n} K(i,j,m,n)} \sum_{m,n} \frac{K(i,j,m,n)I(m,n)}{\sum_{ii,jj} K(ii,jj,m,n)f^{(p)}(ii,jj)}. \quad (9)$$

where $f^{(p)}(i,j)$ is the estimated element concentration after iterating p times, $I(m,n)$ represents the projection value. $K(i,j,m,n)$ denotes the contribution of pixel $(i,j)$ to $I(m,n)$.

$$K(i,j,m,n) = K'(i,j,m,n)f(\theta,s,t)g(\theta,s,t). \quad (10)$$

Where $K'(i,j,m,n)$ is the weighting function without regard to the self-absorption effect. $f(\theta,s,t)$ represents the attenuation of the intensity of the beam when transmitting to the motivated pixel.

$$f(\theta,s,t) = \exp(-\int_{-\infty}^{s} \mu(s',t,E_I)\,ds') \,. \quad (11)$$

$g(\theta,s,t)$ is fluorescence attenuation ratio when transmitting from motivated pixel to the detector.

$$g(\theta,s,t) = \int_{\Omega_0}^{\Omega_1} exp\left(-\int_0^L \mu(s,t,E_f)\,dl\right)d\Omega. \quad (12)$$

### 2.3 Geant4 simulation

Geant4 is a toolkit to simulate the passage of particles through the matter[14],[15]. A large number of experiments and projects use it in a variety of application domains, such as high energy physics, astrophysics and space science, medical physics and radiation protection. The geometry of the system, primary particle and the physical process must be defined to simulate a system. Geant4 provides many physical models such as standard electromagnetic models, low energy electromagnetic models and so on. The standard electromagnetic models are suitable for most of the Geant4 simulation involving electromagnetic process except the low energy particles[16]. Livermore and Penelop are the low energy electromagnetic models to simulate the process about low energy particles. The Low Energy Electromagnetic package provides all kinds of models describing the electromagnetic processes of electrons and positrons, photons, charged hadrons and ions with an eye to detailed features, such as atomic shell effects and charge dependence[17].

XFCT system is simulated by Geant4, and the geometry of the simulation system is shown in figure 1. The far left is a x-ray tube, the right of which is Be-filter. The sample is in the middle, of which the component is shown in table 1. The above of the sample is an energy dispersive detector and the right of the sample is another detector. The environment of the system is vacuum, which is to avoid the attenuation due to the air. X-ray source has two models in the simulation system. One model is the monochromatic x-ray, which is to simulate synchrotron radiation light. The other is the polychrome x-ray, which is x-ray tube model.

To test the veracity of the attenuation correction method, a simulation experiment is performed about cylindrical sample of which the chemical composition is described in table I. As shown in figure 2, the cylindrical sample is composed of a cylinder-shaped tube whose inner diameter is 0.02-mm and external diameter is 0.04-mm and a cylinder whose diameter is 0.02-mm. The angle of each rotation is 5-degree until rotate 180-degree. The number of scanning beam is 81, and the region of scanning is -0.05-mm~0.05-mm. The energy dispersive detector is used to detect the fluorescence through the attenuation of the sample at each projection. The data is stored in two-dimensional matrix, in which the first dimensional is the number of the x ray beam and the second dimensional is the energy of the ray.

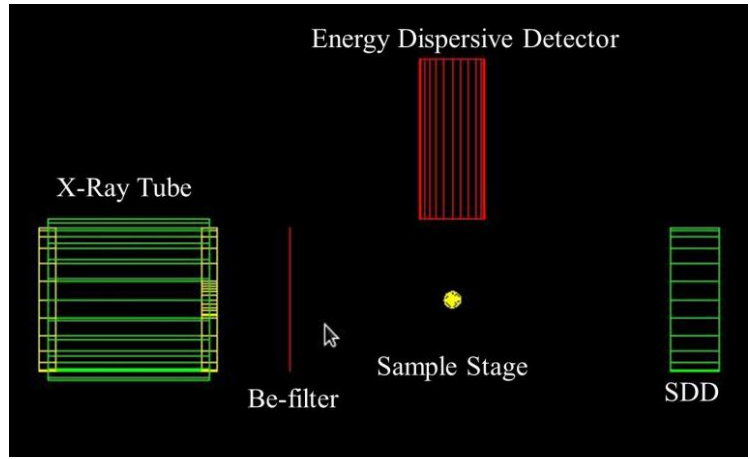

Fig. 1.  The geometry of the simulation X-ray fluorescence computed tomography with Geant4

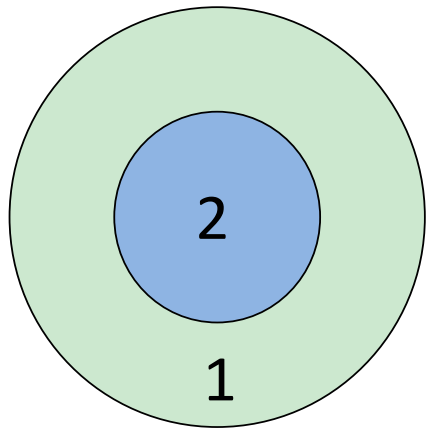

Fig. 2.   The geometry of cylindrical sample

Table 1.  Chemical composition of cylindrical sample

| Phase index | Density(g/cm$^3$) | oxides | Concentration(wt%) |
|---|---|---|---|
| 1 | 2.2 | $SiO_2$ | 100 |
| 2 | 2.4 | $SiO_2$ | 51 |
|  |  | CaO | 20 |
|  |  | $Fe_2O_3$ | 15 |
|  |  | BaO | 14 |

## 3  Result and discussion

Geant4-simulated XFCT images reconstructed with FBP, MLEM without and with attenuation correction for the sample are shown in figure 3. XFCT images reconstructed with FBP were much noisier and sharper than that with MLEM without attenuation correction, which proved that the FBP was less anti-noise than MLEM. On the contrary, the consumed time for MLEM is much longer than that for FBP. For 40*40 pixels, the consumed time of image reconstructed with 10-iterations of MLEM without attenuation correction is about 20-minutes, while the consumed time of image reconstructed with FBP is less than 1-minutes.

In the XFCT images, bottom and central object Fe concentrations were underestimated by 78.2% and 69.8% in FBP, and in MLEM without attenuation correction the ratio is 24.6% and 54.5%. After attenuation correction, Fe concentrations were within 6.3% of true values. Attenuation played a more important role in Ba concentrations image reconstruction. Bottom object Ba concentrations were underestimated by 90.5% and 76.3% in FBP and MLEM without attenuation correction. With MLEM with attenuation correction, Ba concentration was within 14.1% of true values.

Magdalena Bazalova researched the method of getting Pt distribution of Cisplatin, and found that the sensitivity of the K-shell XFCT lowers than that of L-shell XFCT. From now on, however, references about L-shell attenuation correction were not found. This paper presented an attenuation correction method for L-shell of XFCT. The relative error reduced from 88.8% to 14.1% after attenuation correction in Ba concentrations reconstruction. This method can be used to reconstructing L-shell XFCT image of biological samples. The inconvenient of this method is that samples should be irradiated by two different energy x-ray, which increase the radiation dose of samples. Another point we should point that this study simulated an XFCT system using monoenergetic beams, which greatly reduced the complexity in attenuation. For the XFCT system, an array of crystals with individual readout channels, which is just like many CT system, is more practical for experimental XFCT.

## 4 Conclusion

Unknown attenuation maps are expressed by the known quantities and the unknown elemental concentration of interest in attenuation correction of L-shell XFCT, which is proved to be feasible by Geant4 simulation. Results indicate that this method can increase accuracy of the reconstructed XFCT image obviously. This method may have even more application in XFCT attenuation correction of large biological samples.

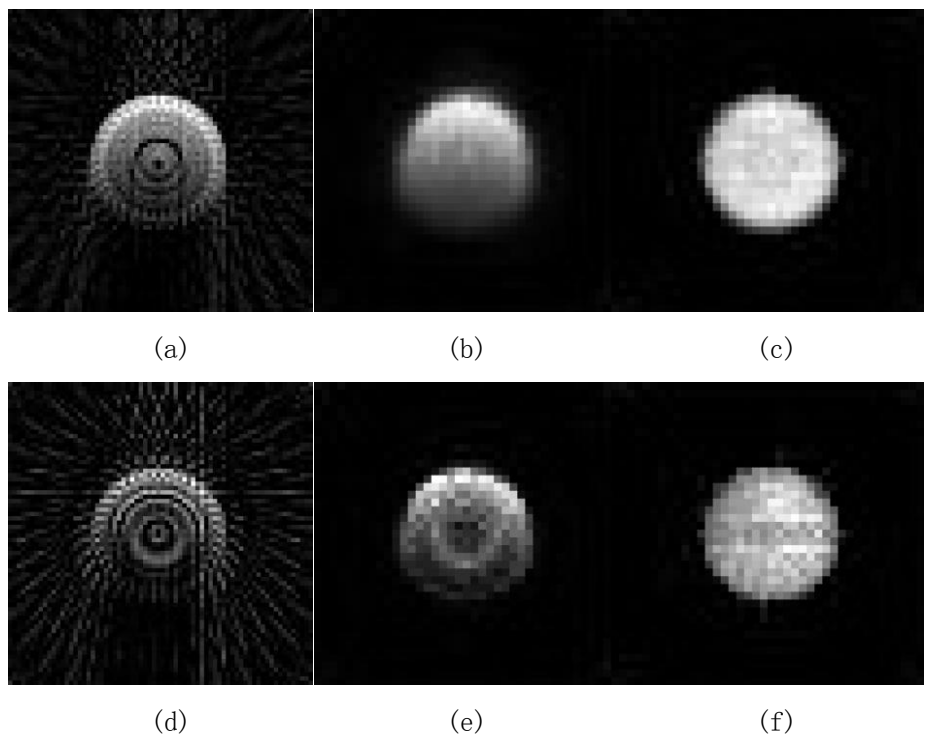

(a) (b) (c)

(d) (e) (f)

Fig. 3. Reconstruction images of Fe distribution and Ba distribution. (a) Fe distribution with FBP. (b) Fe distribution with MLEM without attenuation. (c) Fe distribution with MLEM with attenuation. (d) Ba distribution with FBP. (e) Ba distribution with MLEM without attenuation. (f) Ba distribution with MLEM with attenuation.

Table 2. True and estimated Fe concentration

| Method | Location | True value (g/cm$^3$) | Estimated value (g/cm3) | Difference (%) |
|---|---|---|---|---|
| FBP | Bottom object | 0.252 | 0.055 | 78.2 |
|  | Central object | 0.252 | 0.076 | 69.8 |
| MLEM without attenuation | Bottom object | 0.252 | 0.190 | 24.6 |
|  | Central object | 0.252 | 0.115 | 54.4 |
| MLEM with attenuation | Bottom object | 0.252 | 0.255 | 1.2 |
|  | Central object | 0.252 | 0.236 | 6.3 |

Table 3. True and estimated Ba concentration

| Method | Location | True value (g/cm$^3$) | Estimated value (g/cm3) | Difference (%) |
|---|---|---|---|---|
| FBP | Bottom object | 0.304 | 0.029 | 90.5 |
|  | Central object | 0.304 | 0.035 | 88.5 |
| MLEM without attenuation | Bottom object | 0.304 | 0.072 | 76.3 |
|  | Central object | 0.304 | 0.034 | 88.8 |
| MLEM with attenuation | Bottom object | 0.304 | 0.267 | 12.2 |
|  | Central object | 0.304 | 0.261 | 14.1 |